\documentclass[12pt]{article}
\usepackage{amssymb}
\usepackage{epsfig}
\usepackage{amssymb,amsmath}
\usepackage{slashed}
\usepackage{enumerate}

\newcommand{\bea}{\begin{eqnarray}}
\newcommand{\eea}{\end{eqnarray}}

 \makeatletter
\def\alt{\mathrel{\mathpalette\gl@align<}}
\def\agt{\mathrel{\mathpalette\gl@align>}}
\def\gl@align#1#2{\lower.6ex\vbox{\baselineskip\z@skip\lineskip\z@
\ialign{$\m@th#1\hfil##\hfil$\crcr#2\crcr\sim\crcr}}} \makeatother

\newcommand{\tr}{\,\text{Tr}\,}

\begin{document}
\begin{flushright}
BA-08-17\\
\end{flushright}
\vspace*{1.0cm}

\begin{center}
\baselineskip 20pt {\Large\bf GUT Inflation and Proton Decay after WMAP5} \vspace{1cm}

{\large Mansoor Ur Rehman, Qaisar Shafi, Joshua R. Wickman} \vspace{.5cm}

{\baselineskip 20pt \it 
Bartol Research Institute, Department of Physics and Astronomy, \\
University of Delaware, Newark, DE 19716, USA \\

 }
\vspace{.5cm}

\end{center}

\begin{abstract}

We employ Coleman-Weinberg and Higgs potentials to implement inflation in non-supersymmetric grand unified theories (GUTs) such as SU(5) and SO(10). To realize a scalar spectral index close to 0.96, as indicated by the most recent WMAP 5-year anlaysis, the energy scale of observable inflation turns out to be of order $10^{16}$ GeV. This implies a GUT symmetry breaking scale of similar magnitude, and proton lifetime of order $10^{34}$-$10^{38}$ years. In some SO(10) models with axion dark matter, the scalar leptoquark boson exchange leads to proton decay with a lifetime of order $10^{34}$-$10^{35}$ years.
\end{abstract}

\date{}

\section*{Introduction}

An inflationary scenario \cite{Linde:2007fr} may be termed successful if the following conditions are satisfied:

\vspace{0.3cm}
\begin{enumerate}[1)]
\item The number of e-foldings is sufficient to resolve the horizon and flatness problems.
\item The predicted temperature anisotropies, scalar spectral index $n_s$, tensor-to-scalar ratio 
$r$, etc., are in agreement with the observations.
\item Following  inflation,  there exists a mechanism for explaining the observed  baryon asymmetry.
\item The number density of superheavy magnetic monopoles in GUTs is suitably suppressed to agree
with the upper bounds on their primordial abundance.
\item The model should offer a plausible cold dark matter candidate.
\end{enumerate}
\vspace{0.3cm}

It was shown a long time ago \cite{Shafi:1983bd,Pi:1984pv,Lazarides:1984pq} that non-supersymmetric GUT inflation can be realized by employing a Coleman-Weinberg (CW) type potential for a scalar inflaton field which must be a GUT singlet. The singlet condition ensures that radiative corrections from the gauge sector do not spoil the desired inflationary potential. The monopole problem is resolved by requiring that the GUT symmetry breaking, associated with monopole production, takes place during inflation.  A suitable coupling between the GUT symmetry breaking field $\Sigma$ and the inflaton field $\phi$ is thus needed. This coupling also plays a crucial role in generating an appropriate CW inflationary potential for $\phi$.

The most recent WMAP 5-year \cite{Komatsu:2008hk} analyses seem to favor a scalar spectral index centered around 0.96. Within the CW framework this favors an energy scale for inflation which is close to (1-2)$\times 10^{16}$ GeV \cite{Shafi:2006cs}. This result, as we will show here, has far reaching implications for the symmetry breaking scale of the underlying GUT. In our CW examples below, primarily based on SO(10), the superheavy gauge bosons which mediate proton decay are typically a factor 2-4 larger in mass than the energy scale of inflation preferred by WMAP5. To phrase things somewhat differently, while the most recent proton lifetime limits from Super-K ($\tau (p \rightarrow \pi^0 e^+ )> 8.2 \times 10^{33}$ years \cite{Shiozawa}) require that $M_X \gtrsim 4 \times 10^{15}$ GeV \cite{Nath:2006ut}, the limits on $M_X$  from WMAP seem to favor a value which is up to a factor 10 larger. We thus estimate a proton lifetime of around $10^{34}$-$10^{38}$ years. Proton lifetime estimates of order $10^{34}$ years are achieved in some SO(10)
axion models which contain scalar leptoquarks (3,1,-2/3) with an intermediate mass of order $10^{12}$ GeV.

As an alternative to the CW potential, we will also discuss inflation with a Higgs potential \cite{Kallosh:2007wm,Linde:2007fr}, as well as implications for proton decay. As we will see, such a model predicts values for the scalar spectral index $n_s$, the tensor-to-scalar ratio $r$, the scale of vacuum energy and the proton lifetime that are quite similar to those predicted by the CW model.

\section*{GUT Inflation with a Coleman-Weinberg (CW) Potential}

To simplify the presentation we begin with a brief summary based on SU(5), following references \cite{Shafi:1983bd} and \cite{Linde:2005ht}. The model contains an SU(5) singlet (real) scalar $\phi$ which develops a CW potential from its weak couplings to the adjoint and fundamental Higgs fields $\Sigma$ and $H_5$. The tree-level scalar potential is given by

\begin{eqnarray}
V (\phi,\Sigma,H_5) &=& \frac{1}{4}\,a\, (\tr\Sigma^2)^2+\frac{1}{2}\,b\,\tr \Sigma^4
-\alpha \, ( H_5^\dagger H_5)\tr\Sigma^2+
\frac{\beta}{4} \, (H_5^\dagger H_5)^2\nonumber \\
&+&\gamma \, H_5^\dagger \, \Sigma^2 \, H_5+
\frac{\lambda_1}{4}\phi^4
-\frac{\lambda_2}{2} \phi^2 \tr\Sigma^2
+\frac{\lambda_3}{2} \phi^2\,H_5^\dagger \, H_5.
\label{VSU5}
\end{eqnarray}
The coefficients $a$, $b$, $\alpha$ and $\beta$ are taken to be of order $g^2$, so that most radiative corrections in the ($\Sigma$, $H_5$) sector can be neglected. We assume a somewhat smaller value for the coefficient $\gamma$ and 
take $0<\lambda_i\ll g^2$ and $\lambda_1\lesssim \text{max}(\lambda_2^2,\lambda_3^2)$.

Radiative corrections due to the couplings $\phi^2 \tr\Sigma^2$ and $ \phi^2\,H_5^\dagger \,H_5$ induce a Coleman-Weinberg potential for $\phi$, which is given by

\begin{eqnarray}
V(\phi)=A\,\phi^4\,\left(\ln \left( \frac{\phi}{M} \right) +C\right)+ V_0,
\end{eqnarray}
with 
\begin{eqnarray}
A=\frac{\lambda_2^2}{16\,\pi^2}\,
\left(1+\frac{25\,g^4}{16\,\lambda_c^2}+
\frac{14\,b^2}{9\,\lambda_c^2}\right),
\label{CWA}
\end{eqnarray}
and $\displaystyle \lambda_c=a+\frac{7}{15}\,b$. The SU(5) symmetry is broken to SU(3)$_c$ $\times$ SU(2)$_L$ 
$\times$ U(1)$_Y$ when $\Sigma$ acquires a vacuum expectation value (VEV)

\begin{eqnarray}
\langle \Sigma \rangle =\sqrt{\frac{1}{15}}\,\sigma\cdot
\mbox{diag}\left(1,1,1,-\frac{3}{2},-\frac{3}{2}\right),
\label{vev}
\end{eqnarray}
where
\begin{eqnarray}
\sigma^2=\frac{2\,\lambda_2}{\lambda_c}\,\phi^2.
\label{vev2}
\end{eqnarray}
With $\Sigma$ given in Eq. (\ref{vev}), the $(\phi , \sigma)$ sector of the 
effective potential can be written as

\begin{equation}
V=\frac{\lambda_c}{16}\,\sigma^4-\frac{\lambda_2}{4}\,\sigma^2\,\phi^2+A\,\phi^4\,\left(\ln \left(  \frac{\phi}{{M} }  \right) + C\right)+ V_0.
\end{equation}
Using Eq. (\ref{vev2}) and an appropriate choice of the normalization constant $C$, the effective potential for the inflaton field $\phi$ can be expressed in the standard form \cite{Linde:2005ht}

\begin{equation}
V(\phi)=A\,\phi^4\,\left[\ln\left(\frac{\phi}{M}\right)-\frac{1}{4}\right]+\frac{A\,M^4}{4}\ ,
\label{CWP}
\end{equation}
where $M = \langle \phi \rangle$ denotes the VEV of $\phi$ at the minimum. Note that the vacuum energy density at the origin is given by $V_0 = \frac{A M^4}{4}$ such that $V(\phi=M)=0$, and the corresponding minimum in $\sigma$ is located at $\sigma_0=\sqrt{\frac{2\,\lambda_2}{\lambda_c}}\,M$. The mass of the superheavy gauge bosons $X$ which mediate proton decay is given by

\begin{equation}
M_X=\sqrt{\frac{5}{3}}\,\frac{g\,\sigma_0}{2}=\sqrt{\frac{5 \, \lambda_2 \, g^2}{3\lambda_c \, A^{1/2}}}\, V_0^{1/4}.
\end{equation}
Thus $M_X$ is estimated to be a factor 2-4 larger than the scale of vacuum energy during inflation.
This is to be contrasted with the simplest supersymmetric hybrid inflation models  in which 
the corresponding $M_X$ can easily be 1-2 orders of magnitude larger than the scale of vacuum energy during inflation \cite{Dvali:1994ms}.

The inflationary potential in Eq. (\ref{CWP}) is typical for the new inflation scenario \cite{Linde:1981mu}, where inflation takes place near the maximum. However, as shown in reference \cite{Shafi:2006cs}, depending on the value of $V_0$, the inflaton can have small or large values compared to the Planck scale during observable inflation. In the latter case, observable inflation takes place near the minimum of $V(\phi)$, and the model mimics chaotic inflation \cite{Linde:1983gd}. Indeed, we will see from Table \ref{tableI} that in order to obtain a scalar spectral index close to 0.96, the energy scale for observable inflation is typically on the order of $10^{16}$ GeV. 

So far, we have been discussing the case in which $\phi <M$ during inflation.  Alternatively, the inflaton may roll toward its minimum starting from values larger than the VEV, similar to simple models of chaotic inflation \cite{Kallosh:2007wm}. This is true in both of the inflationary models (CW and Higgs) that we consider here.  For shorthand, we henceforth denote these regimes as the BV (below VEV) and AV (above VEV) branches.

\subsection*{Results for the CW Model}

The inflationary slow-roll parameters are defined as \cite{Liddle:1992wi}
\begin{equation}
\epsilon =
 \frac{1}{2}\left(\frac{V'}{V}\right)^2 \,,\quad
\eta=\left(\frac{V''}{V}\right) \,,\quad
\xi^2=\left(\frac{V'\, V'''}{V^2}\right) \,.
\end{equation}
(Here and below we use units $m_P=1$, where $m_P\simeq2.4\times10^{18}$ GeV is the reduced Planck mass.) The slow-roll approximation is valid as long as the conditions $\epsilon \ll 1$ and $\eta \ll 1$ hold. In this case the spectral index $n_{s}$, the tensor-to-scalar ratio $r$ and the running of the spectral index $\frac{d n_{s}}{d \ln k}$ are given by
\begin{eqnarray}
n_{s} \!&\simeq&\! 1 - 6 \epsilon + 2 \eta \label{ns}\\
r \!&\simeq&\! 16 \epsilon \\
\frac{d n_{s}}{d \ln k} \!\!&\simeq&\!\!
16 \epsilon \eta - 24 \epsilon^2 - 2 \xi^2.
\end{eqnarray}
The number of e-foldings after the comoving scale $l_0=2\pi/k_0$ has crossed the horizon is
given by
\begin{equation} \label{efold1}
N_0=\frac{1}{2}\int^{\phi_0}_{\phi_e}\frac{H(\phi)\rm{d}\phi}{H'(\phi)}, \end{equation}
where $\phi_0$ is the value of the field when the scale corresponding to $k_0$ exits the horizon, and $\phi_e$ is the value of the field at the end of inflation. The value of $\phi_e$ is given by the condition $2(H'(\phi_e)/H(\phi_e))^2=1$, which can be calculated from the Hamilton-Jacobi equation \cite{Salopek:1990jq}
\begin{equation}
[H'(\phi)]^2-\frac{3}{2}H^2(\phi)=-\frac{1}{2}V(\phi)\,.
\end{equation}
The amplitude of the curvature perturbation $\Delta_{\mathcal{R}}$ is given by
\begin{equation} \label{perturb}
\Delta_{\mathcal{R}}=\frac{1}{2\sqrt{3}\pi }\frac{V^{3/2}}{|V'|}\,.
\end{equation}
(Note that, for added precision, we include in our calculations the first order corrections in the slow-roll expansion for the quantities $n_s$, $r$, and $\Delta_{\mathcal{R}}$ \cite{Stewart:1993bc}.)

\begin{table}[t]
{\centering
\resizebox{!}{5.1 cm}{
\begin{tabular}{||c|c|c|c|c|c|c|c|c|c||}
\hline
 $V^{1/4}_0$(GeV) & $V(\phi_0)^{1/4}$(GeV) & A ($10^{-14}$)&  $M$ & $\phi_0$ & $\phi_e$ & $n_s$ & $r$ & $T_r$(GeV) & $\frac{dn_{s}}{d\ln k}(-10^{-3})$   \\
\hline \hline
$ 2.\times 10^{15}$ & $2.00\times 10^{15}$ & 3.7 & 2.65 & 0.11 & 2.05 & 0.9369 & 0.000013 & $1.54\times 10^8$ & 1.31 \\
\hline
$ 5.\times 10^{15}$ & $5.00\times 10^{15}$ & 6.1 & 5.85 & 0.65 & 5.06 & 0.9375 & 0.000496 & $4.14\times 10^7$ & 1.21 \\
\hline
$ 8.\times 10^{15}$ & $7.98\times 10^{15}$ & 7.3 & 8.95 & 1.71 & 8.09 & 0.9421 & 0.00322 & $2.09\times 10^7$ & 0.982 \\
\hline
$ 1.\times 10^{16}$ & $9.93\times 10^{15}$ & 7.2 & 11.2 & 2.82 & 10.3 & 0.9465 & 0.00773 & $1.49\times 10^7$ & 0.832 \\
\hline
$ 1.25\times 10^{16}$ & $1.23\times 10^{16}$ & 6.1 & 14.6 & 4.91 & 13.7 & 0.9525 & 0.0180 & $1.05\times 10^7$ & 0.703 \\
\hline
$ 1.5\times 10^{16}$ & $1.44\times 10^{16}$ & 4.4 & 19.0 & 8.18 & 18.1 & 0.9578 & 0.0341 & $7.65\times 10^6$ & 0.652 \\
\hline
$ 1.75\times 10^{16}$ & $1.61\times 10^{16}$ & 2.8 & 24.8 & 13.1 & 23.9 & 0.9613 & 0.0538 & $5.73\times 10^6$ & 0.654 \\
\hline
$ 2.\times 10^{16}$ & $1.74\times 10^{16}$ & 1.7 & 32.3 & 19.9 & 31.4 & 0.9630 & 0.0730 & $4.40\times 10^6$ & 0.670 \\
\hline
$ 2.25\times 10^{16}$ & $1.83\times 10^{16}$ & 0.99 & 41.4 & 28.5 & 40.5 & 0.9637 & 0.0889 & $3.45\times 10^6$ & 0.686 \\
\hline
$ 2.5\times 10^{16}$ & $1.89\times 10^{16}$ & 0.61 & 52.0 & 38.8 & 51.1 & 0.9638 & 0.101 & $2.77\times 10^6$ & 0.697 \\
\hline
$ 2.75\times 10^{16}$ & $1.93\times 10^{16}$ & 0.39 & 64.0 & 50.5 & 63.0 & 0.9636 & 0.110 & $2.27\times 10^6$ & 0.704 \\
\hline
$ 3.\times 10^{16}$ & $1.96\times 10^{16}$ & 0.26 & 77.1 & 63.5 & 76.2 & 0.9635 & 0.117 & $1.90\times 10^6$ & 0.710 \\
\hline
$ 4.\times 10^{16}$ & $2.03\times 10^{16}$ & 0.072 & 142 & 128 & 141 & 0.9628 & 0.133 & $1.05\times 10^6$ & 0.724 \\
\hline \hline 
$ 4.\times 10^{16}$ & $2.14\times 10^{16}$ & 0.047 & 157 & 172 & 159 & 0.9607 & 0.165 & $1.0\times 10^6$ & 0.743 \\
\hline
$ 3.\times 10^{16}$ & $2.17\times 10^{16}$ & 0.13 & 92.5 & 108 & 93.5 & 0.9600 & 0.174 & $1.73\times 10^6$ & 0.744 \\
\hline
$ 2.75\times 10^{16}$ & $2.19\times 10^{16}$ & 0.17 & 79.1 & 94.3 & 80.1 & 0.9597 & 0.178 & $2.04\times 10^6$ & 0.745 \\
\hline
$ 2.5\times 10^{16}$ & $2.20\times 10^{16}$ & 0.22 & 66.9 & 82.2 & 67.9 & 0.9592 & 0.183 & $2.45\times 10^6$ & 0.747 \\
\hline
$ 2.25\times 10^{16}$ & $2.22\times 10^{16}$ & 0.3 & 55.8 & 71.3 & 56.8 & 0.9587 & 0.189 & $2.98\times 10^6$ & 0.750 \\
\hline
$ 2.\times 10^{16}$ & $2.24\times 10^{16}$ & 0.41 & 45.8 & 61.5 & 46.9 & 0.9580 & 0.196 & $3.69\times 10^6$ & 0.755 \\
\hline
$ 1.75\times 10^{16}$ & $2.26\times 10^{16}$ & 0.57 & 37.0 & 52.9 & 38.0 & 0.9570 & 0.205 & $4.70\times 10^6$ & 0.763 \\
\hline
$ 1.5\times 10^{16}$ & $2.30\times 10^{16}$ & 0.79 & 29.2 & 45.4 & 30.3 & 0.9557 & 0.216 & $6.17\times 10^6$ & 0.775 \\
\hline
$ 1.25\times 10^{16}$ & $2.33\times 10^{16}$ & 1.1 & 22.5 & 39.1 & 23.5 & 0.9539 & 0.230 & $8.44\times 10^6$ & 0.794 \\
\hline
\end{tabular} 
\par} 
\caption{Predicted values of various inflationary parameters using the Coleman-Weinberg potential given in Eq. (\ref{CWP}). Here we show only those values which fall inside the WMAP5 $2\sigma$ bounds (see Fig. \ref{combo_nsr}), and also omit values with very low reheat temperature $T_{r}(< 10^6$ GeV). Note that unless otherwise specified, we use units of $m_P=1$.
Although $M$, $\phi_0$ and $\phi_e$ carry transplanckian values, the vacuum energy scale during observable inflation
is well below $m_P$.} \label{tableI}
\centering}
\end{table}

To calculate the magnitude of $A$ and the inflationary parameters, we use these standard equations above. The WMAP5 value for $\Delta_{\mathcal{R}}$ is $4.91\times10^{-5}$ for $k_0 = 0.002$ Mpc$^{-1}$ \cite{Komatsu:2008hk}. The observable number of e-foldings corresponding to the same scale is
\begin{equation}
N_{0}\simeq 53+\frac{2}{3}\ln\left[\frac{V(\phi_0)^{1/4}}{10^{15}\rm{ GeV}}\right] +\frac{1}{3}\ln\left[\frac{T_r}{10^{9}\rm{ GeV}}\right] .
\end{equation}
For our calculations here and in the next section, we use the formula given in Eq. (\ref{TR}) to obtain values for the reheat temperature $T_{r}$. This formula is based on an SO(10) GUT model, and yields values on the order of $10^6$-$10^7$ GeV. In contrast, SU(5) considerations give $T_{r}\sim10^9$ GeV.

\begin{figure}[t] 
\begin{center}
\includegraphics[angle=0, width=13cm]{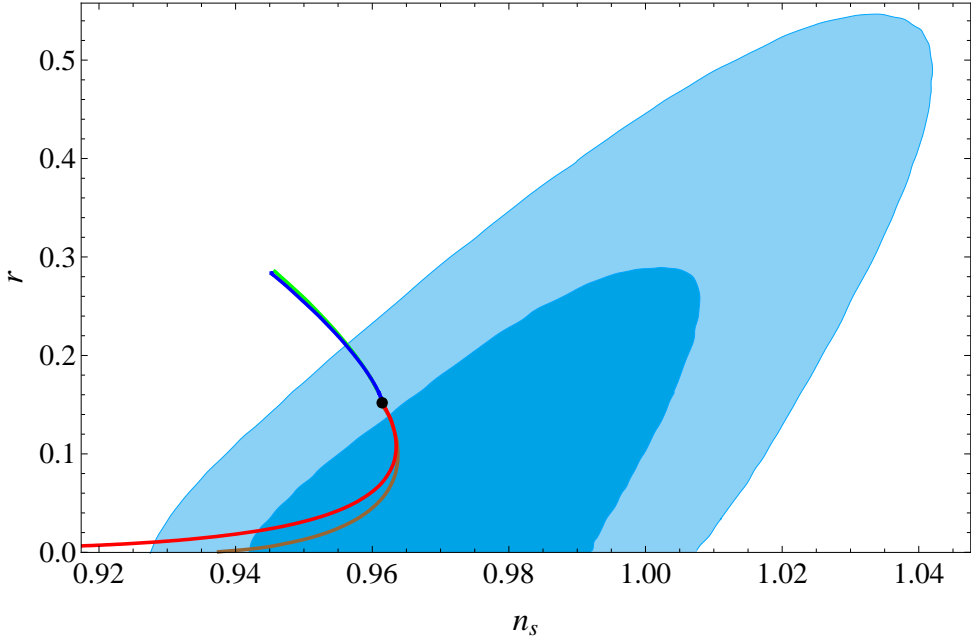} 
\caption{$r$ vs. $n_{s}$ for the Higgs and Coleman-Weinberg models, shown together with the WMAP5 contours (68\% and 95\% confidence levels) \cite{Komatsu:2008hk}. In each model, inflation is allowed both below the VEV (BV) and above the VEV (AV). The BV and AV branches for the Higgs (CW) potential are shown in red (brown) and blue (green), respectively. The black circle corresponds to a quadratic potential with $N_0 = 52$, 
$n_s = 0.961 $ and $r = 0.152 $.} \label{combo_nsr}
\end{center}
\end{figure}

Our predictions for the values of various parameters are displayed in Table \ref{tableI}. As we see in Fig. \ref{combo_nsr}, both the BV and AV branches fall within the WMAP5 $2\sigma$ bounds for some range, and part of the BV branch lies inside the $1\sigma$ bound. As discussed earlier, larger values of $V_0^{1/4}$ result in observable inflation occurring closer to the VEV $M$. For the largest values of $V_0^{1/4}$, the potential is effectively given by $V=\frac{1}{2} m_{\phi}^2 (\Delta\phi)^2$, where $\Delta\phi=M-\phi$ denotes the deviation of the field from the minimum and $m_{\phi} = 2 \sqrt{A} M$  $\left( = \frac{4\sqrt{V_0}}{M} \right) $ is the inflaton mass in the CW model. This well-known monomial model \cite{Linde:1983gd} predicts $m_{\phi}\simeq2\times10^{13}$ GeV, $\Delta\phi_0\simeq2\sqrt{N_0}$, $n_s\simeq 1-\frac{2}{N_0}$, and $r\simeq 4(1-n_s)$, corresponding to $V(\phi_0)\simeq(2\times10^{16}$ GeV$)^4$.  This is the region in which the two branches meet, i.e. both the BV and AV branches converge to quadratic inflation in the high-$V_0$ limit.

In the opposite limit with lower values of $V_0$, the BV branch reduces to new inflation models with an effective potential $V=V_0 \left( 1-\vert \lambda_{\phi}\vert(\frac{\phi}{M})^4 \right) $, where $\lambda_{\phi}=\text{ln}\left( \frac{\phi}{M}\right)^4 $, leading to $n_s\simeq1-\frac{3}{N_0}$ and $r\simeq \frac{16}{3}(1-n_s)\vert \lambda_{\phi_0}\vert \left( \frac{\phi_0}{M}\right)^4 \approx 0$. On the other hand, the AV branch is asymptotic to quartic inflation with an effective potential $V = \left( \frac{V_0}{M^4} \right)  \lambda_{\phi}\,\phi^4$, resulting in $n_s\simeq 1-\frac{3}{N_0}$ and $r\simeq \frac{16}{3}(1-n_s)$.

\begin{figure}[t] 
\begin{center}
\includegraphics[angle=0, width=13cm]{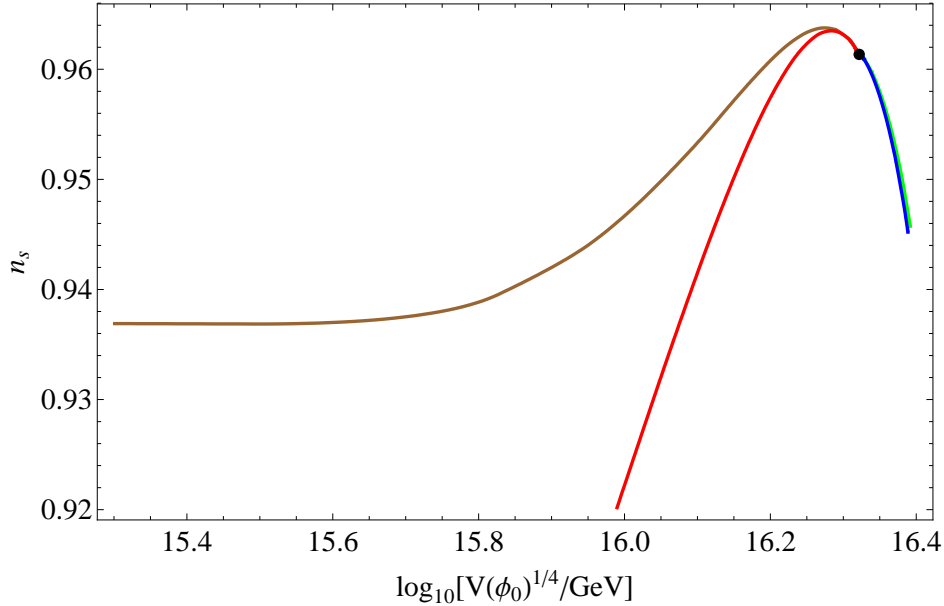} 
\caption{$n_{s}$ vs. $\text{log}_{10}[V(\phi_0)^{1/4}/\text{GeV}]$ for the CW 
and Higgs models. The BV and AV branches for the Higgs (CW) potential are shown in red (brown) and blue 
(green), respectively. The black circle corresponds to a quadratic potential with $N_0 = 52$ and 
$n_s = 0.961 $.} \label{nsV}
\end{center}
\end{figure}

As previously indicated, Table \ref{tableI} shows that the energy scale of observable inflation in this model is $\sim 10^{16}$ GeV.  It is instructive to see how $n_{s}$ changes with this quantity, and this is displayed in Fig. \ref{nsV}.

\section*{GUT Inflation with a Higgs Potential}

In this section we implement inflation by employing a Higgs potential given by \cite{Kallosh:2007wm,Destri:2007pv,Smith:2008pf}
\begin{equation}
V\left(\phi \right) =V_{0}-\mu^{2}\phi^{2}+\lambda\phi^{4}.
\end{equation}
It is useful to rewrite this in terms of the vacuum potential $V_{0}$ and the vacuum expectation value $M$ of the inflaton \cite{Kallosh:2007wm,Smith:2008pf}:
\begin{equation}
V\left(\phi \right)=V_{0}\left[ 1-\left( \frac{\phi}{M}\right) ^{2}\right] ^{2}.
\label{Higgs}
\end{equation}

As mentioned earlier, and previously discussed in reference \cite{Kallosh:2007wm}, inflation may occur above or below the VEV  $M$. We proceed in the same way as before, noting that $V_{0}$ is not related to other parameters in a simple way as it was in the CW case. In this model, the inflaton mass is given by $m_{\phi}=\frac{2\sqrt{2V_{0}}}{M}$.

\begin{table}[t]
{\centering
\resizebox{!}{4.8 cm}{
\begin{tabular}{||c|c|c|c|c|c|c|c|c||}
\hline
 $V^{1/4}_0$ (GeV) & $V(\phi_0)^{1/4}$ (GeV) & $M$ & $\phi_0$ & $\phi_e$ & $n_s$ & $r$ & $T_{r}$ (GeV) & $\frac{dn_{s}}{d\ln k} (-10^{-3})$ \\
\hline\hline
 $1.19\times 10^{16}$ & $1.18\times 10^{16}$ & 11.57 & 1.482 & 10.63 & 0.9360 & 0.0151 & $1.47\times 10^7$ & 0.316 \\
 \hline
 $1.37\times 10^{16}$ & $1.35\times 10^{16}$ & 13.26 & 2.482 & 12.32 & 0.9466 & 0.0259 & $1.19\times 10^7$ & 0.427 \\
 \hline
 $1.58\times 10^{16}$ & $1.52\times 10^{16}$ & 15.86 & 4.331 & 14.90 & 0.9550 & 0.0424 & $9.44\times 10^6$ & 0.537 \\
 \hline
 $1.83\times 10^{16}$ & $1.68\times 10^{16}$ & 20.00 & 7.727 & 19.03 & 0.9604 & 0.0638 & $7.28\times 10^6$ & 0.621 \\
 \hline
 $2.11\times 10^{16}$ & $1.81\times 10^{16}$ & 26.41 & 13.50 & 25.43 & 0.9628 & 0.0855 & $5.49\times 10^6$ & 0.669 \\
 \hline
 $2.44\times 10^{16}$ & $1.90\times 10^{16}$ & 35.69 & 22.32 & 34.70 & 0.9635 & 0.103 & $4.09\times 10^6$ & 0.692 \\
 \hline
 $2.89\times 10^{16}$ & $1.97\times 10^{16}$ & 51.57 & 37.83 & 50.58 & 0.9634 & 0.119 & $2.86\times 10^6$ & 0.708 \\
 \hline
 $3.64\times 10^{16}$ & $2.02\times 10^{16}$ & 83.82 & 69.79 & 82.82 & 0.9629 & 0.131 & $1.78\times 10^6$ & 0.719 \\
 \hline
 $4.33\times 10^{16}$ & $2.04\times 10^{16}$ & 119.9 & 105.7 & 118.9 & 0.9626 & 0.137 & $1.25\times 10^6$ & 0.726 \\
 \hline\hline
 $4.33\times 10^{16}$ & $2.12\times 10^{16}$ & 129.5 & 144.2 & 130.5 & 0.9611 & 0.161 & $1.21\times 10^6$ & 0.739 \\
 \hline
 $3.64\times 10^{16}$ & $2.13\times 10^{16}$ & 93.31 & 108.1 & 94.33 & 0.9608 & 0.164 & $1.69\times 10^6$ & 0.738 \\
 \hline
 $2.89\times 10^{16}$ & $2.16\times 10^{16}$ & 60.86 & 75.89 & 61.89 & 0.9603 & 0.171 & $2.63\times 10^6$ & 0.739 \\
 \hline
 $2.44\times 10^{16}$ & $2.18\times 10^{16}$ & 44.60 & 59.82 & 45.63 & 0.9597 & 0.178 & $3.66\times 10^6$ & 0.742 \\
 \hline
 $2.11\times 10^{16}$ & $2.20\times 10^{16}$ & 34.71 & 50.13 & 35.74 & 0.9591 & 0.185 & $4.79\times 10^6$ & 0.745 \\
 \hline
 $1.83\times 10^{16}$ & $2.22\times 10^{16}$ & 27.24 & 42.88 & 28.28 & 0.9583 & 0.192 & $6.24\times 10^6$ & 0.751 \\
 \hline
 $1.58\times 10^{16}$ & $2.24\times 10^{16}$ & 21.58 & 37.46 & 22.63 & 0.9573 & 0.201 & $8.09\times 10^6$ & 0.760 \\
 \hline
 $1.37\times 10^{16}$ & $2.27\times 10^{16}$ & 17.27 & 33.42 & 18.33 & 0.9563 & 0.210 & $1.04\times 10^7$ & 0.771 \\
 \hline
 $1.19\times 10^{16}$ & $2.29\times 10^{16}$ & 13.97 & 30.40 & 15.04 & 0.9551 & 0.219 & $1.34\times 10^7$ & 0.784 \\
 \hline
 $1.03\times 10^{16}$ & $2.32\times 10^{16}$ & 11.40 & 28.12 & 12.48 & 0.9539 & 0.228 & $1.71\times 10^7$ & 0.802 \\
\hline
\end{tabular} }
\par} 
\caption{Predicted values of various inflationary parameters using the Higgs potential given in Eq. (\ref{Higgs}), analogous to Table \ref{tableI}.}
\centering
\label{tableII}
\end{table}

The results of this calculation are given in Table \ref{tableII}. For small values of $V_{0}$, the AV branch approaches quartic inflation with an effective potential $V=\left(\frac{V_0}{M^4} \right)\phi^4$, which is the same as in the CW case
apart from the slowly varying function $\lambda_{\phi}$. On the other hand, the BV branch approaches new inflation with an effective potential $V=V_0\left( 1-2\left(\frac{\phi}{M} \right)^2 \right) $ in a region disfavored by WMAP5. This leads to $n_s\simeq 1-\frac{2}{N_0}\, \text{ln} \left( \frac{M}{\sqrt{2}\phi_0}\right) $ and $r\simeq 8(1-n_s)e^{-N_0(1-n_s)}$. In contrast to the AV regime, the CW and Higgs models produce distinct predictions in the BV regime, which is apparent from Fig. \ref{combo_nsr}. For large $V_{0}$, both branches converge to quadratic inflation and to the CW solution. As in the CW model, observable inflation in this region takes place near the VEV $M$.  Hence, it is not surprising that the two models yield very similar results.

As shown in Fig. \ref{combo_nsr}, the Higgs model yields results compatible with WMAP5 for a wide range of values. Analogous to the CW case, BV inflation appears to be somewhat more favorable than AV inflation, although both branches fall within $2\sigma$ of the central value for some range. In this model, the energy scale of observable inflation is again $\sim (1$-$2)\times 10^{16}$ GeV, except for regions far outside the WMAP bounds. This can be seen in Table \ref{tableII} for the most favored range of $(r,n_{s})$ values, as well as in Fig. \ref{nsV}.


Note that radiative corrections to the inflationary potential, especially from Yukawa interactions, can modify the tree-level inflationary predictions in Table \ref{tableII}. This has recently been discussed for $\phi^2$ and $\phi^4 $ potentials in reference \cite{NeferSenoguz:2008nn}.

\begin{figure}[t] 
\begin{center}
\includegraphics[angle=0, width=13cm]{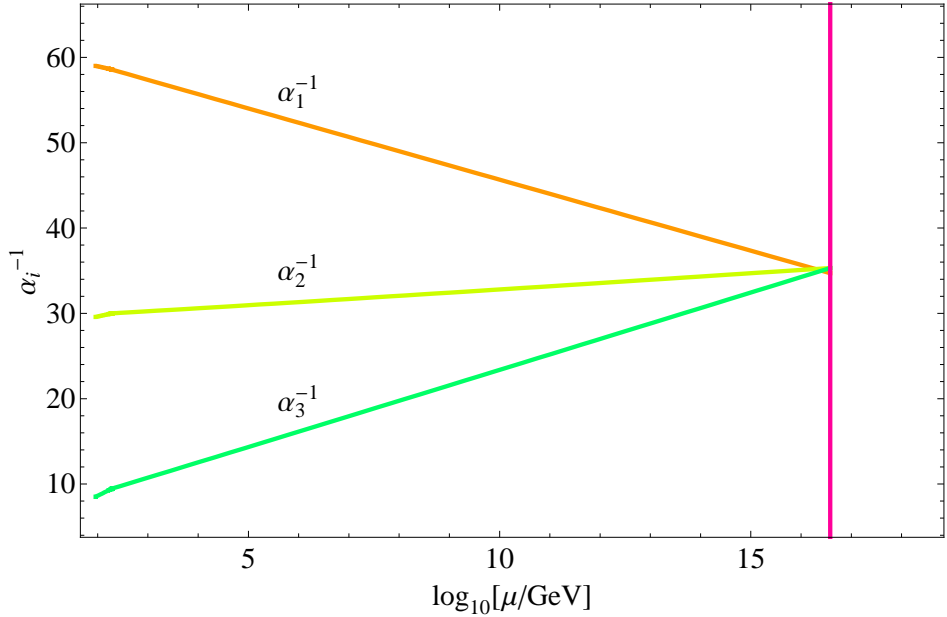} 
\caption{Gauge coupling unification in the SU(5) model with additional fermions 
$Q+\overline{Q}+D_{c}+\overline{D}_{c}$ at mass scale $\sim 200 $ GeV.} \label{SU(5)}
\end{center}
\end{figure}

\section*{Realistic SU(5) and SO(10) Models}

It is well known that satisfactory gauge coupling unification is not achieved in the minimal non-supersymmetric SU(5) model \cite{Dorsner:2005fq}. However, this situation can be improved by introducing additional matter fields at a low energy scale. For example, following reference \cite{Chkareuli:1994ng}, if we introduce vectorlike fermions $Q(3,2,1/6)+\overline{Q}(\overline{3},2,-1/6)$ and $D_{c}(\overline{3},1,1/3)+\overline{D}_{c}( 3,1,-1/3)$ with masses $\sim 200 $ GeV, the gauge couplings unify at a scale $M_X \sim 3.8 \times 10^{16}$ GeV, as shown in Fig. \ref{SU(5)}. Now using $V_0^{1/4}\sim M_X/2 $ and referring to Table \ref{tableI}, we obtain a corresponding value of the spectral index $n_s \sim$ 0.962 which is well inside the 1$\sigma$ bound of WMAP5 (see Fig. \ref{combo_nsr}).

As mentioned earlier an inflationary scenario should provide, among other things, both a suitable cold dark matter candidate and an explanation of the observed baryon asymmetry. We will now consider a class of SO(10) GUT models in which these conditions are readily met \cite{Holman:1982tb}. In these models axions comprise the dark matter of the universe, and the observed baryon asymmetry arises via leptogenesis \cite{Fukugita:1986hr,Lazarides:1991wu}. Clearly, axion dark matter can also be introduced in the SU(5) model.

The breaking of SO(10) to the Standard Model (SM) proceeds via an intermediate step SU(3)$_c$ $\times$ SU(2)$_L$ $\times$ SU(2)$_R$ $\times$ U(1)$_{B-L}$ \cite{Pati:1974yy}. The U(1)$_{PQ}$ (Peccei-Quinn) symmetry and SU(2)$_R \times$U(1)$_{B-L}$ are both spontaneously broken at some scale $M_{B-L}$. Axion dark matter physics requires that $M_{B-L}$ is around $10^{11}$-$10^{12}$ GeV \cite{Preskill:1982cy}. Furthermore, from our earlier discussion, satisfactory CW inflation requires that the SO(10) symmetry breaking scale be close to, indeed somewhat larger than $10^{16}$ GeV. It is intriguing that the scale of $B-L$ and axion symmetry breaking and the inflaton mass are of the same order of magnitude. This enables the inflaton to produce right-handed neutrinos whose subsequent decay produces the observed baryon asymmetry via non-thermal leptogenesis \cite{Lazarides:1991wu}.

The fermion content of the SO(10) model consists of three SM families in the 16-dimensional spinor representations, as well as two fermion matter multiplets in the 10-plet representations \cite{Holman:1982tb}. These 10-plet fields are included in order to resolve the well-known axion domain wall problem \cite{Sikivie:1982qv}, by ensuring that a residual, discrete $PQ$ symmetry coincides with the center, $Z_4$, of SO(10) \cite{Lazarides:1982tw}. Under U(1)$_{PQ}$, the fermion fields transform as follows:

\begin{equation}
\psi_{16}^{(j)}\longrightarrow e^{(i\theta)}\psi_{16}^{(j)}\, (j=1,2,3), \, \,
\psi_{10}^{(\alpha)}\longrightarrow e^{(-2i\theta)}\psi_{10}^{(\alpha)} \, (\alpha = 1,2).
\label{10plet}
\end{equation}
The SO(10) symmetry breaking proceeds as follows:
\begin{eqnarray}
{\rm SO(10)}\times {\rm U(1)}_{PQ}\stackrel{210}{\longrightarrow} {\rm SU(3)}_c\times {\rm SU(2)}_L\times {\rm SU(2)}_R\times {\rm U(1)}_{B-L}\times {\rm U(1)}_{PQ} \nonumber\\
\stackrel{45,126}{\longrightarrow} {\rm SU(3)}_c\times {\rm SU(2)}_L\times {\rm U(1)}_{Y}\stackrel{10}{\rightarrow} {\rm SU(3)}_c\times {\rm U(1)}_{em},
\end{eqnarray}
where the Higgs fields necessary to implement this chain are as indicated. Under
U(1)$_{PQ}$, the Higgs fields transform as follows:
\begin{equation}
\phi^{(210)}\rightarrow \phi^{(210)}, \,\, 
\phi^{(126)}\rightarrow e^{2i\theta}\phi^{(126)}, \,\, 
\phi^{(45)}\rightarrow e^{4i\theta}\phi^{(45)}, \,\, 
\phi^{(10)}\rightarrow e^{-2i\theta}\phi^{(10)}.
\end{equation}
As in the fermion case, these U(1)$_{PQ}$ transformation properties ensure that the action of the residual $PQ$ symmetry on these fields is identical to that of the center of SO(10). Note that all Higgs fields except for $\phi^{(210)}$ are complex.
 The allowed Yukawa couplings are (in schematic form)
\begin{equation}
\psi_{16}\psi_{16}\phi^{(10)}, \, \, \psi_{16}\psi_{16}\phi^{(126)\dagger},
\, \, \psi_{10}\psi_{10}\phi^{(45)}.
\label{fc}
\end{equation}
The allowed Higgs couplings include
\begin{equation}
\phi^{(210)}\phi^{(126)\dagger}\phi^{(126)\dagger}\phi^{(45)}, \, \, 
\phi^{(210)}\phi^{(126)\dagger}\phi^{(10)}\phi^{(45)}, \, \, 
\phi^{(210)}\phi^{(126)}\phi^{(10)}.
\label{hc}
\end{equation}
These couplings guarantee that U(1)$_{PQ}$ is the only global symmetry present. They also guarantee that $\phi^{(45)}\rightarrow -\phi^{(45)}$ is not a summetry of the Langragian so that the domain-wall problems associated with this symmetry can be avoided \cite{Sikivie:1982qv}. Now, $\langle \phi^{(210)}\rangle$ cannot break U(1)$_{PQ}$ since it is neutral under U(1)$_{PQ}$. Hence U(1)$_{PQ}$ is broken at the intermediate scale $M_{B-L}$ by $\langle \phi^{(126)}\rangle$, $\langle \phi^{(45)}\rangle$. 

To make contact with the discussion on CW-based SU(5) inflation, we need to make sure that the tree-level couplings in Eqs.(\ref{fc}) and (\ref{hc}) respect scale invariance. In particular, any cubic scalar coupling which respects SO(10)$\times$U(1)$_{PQ}$ is accompanied by the inflaton field $\phi$ and suitable dimensionless coefficients.

\begin{figure}[t] 
\begin{center}
\includegraphics[angle=0, width=13cm]{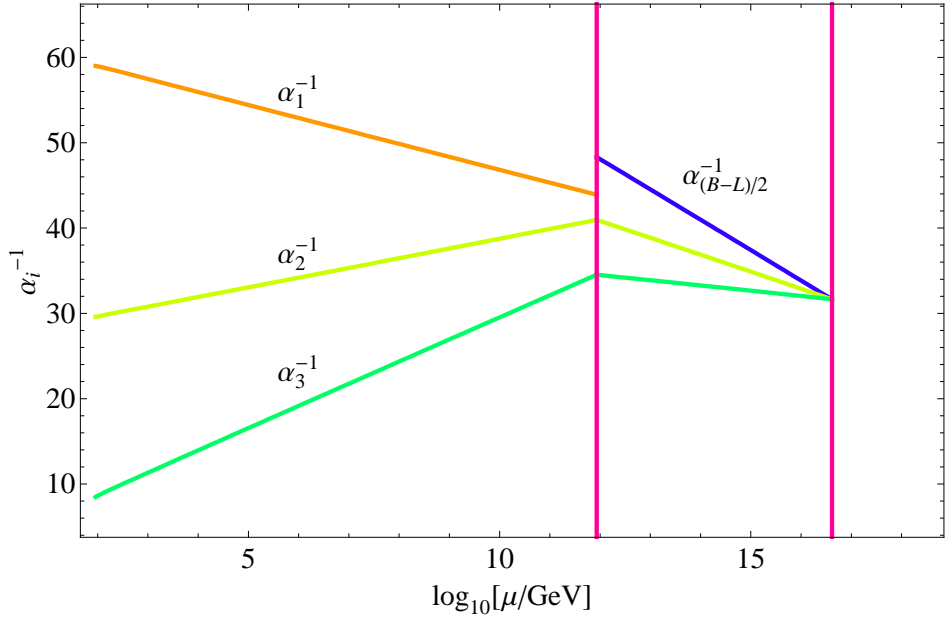} 
\caption{Gauge coupling unification in the SO(10) axion model at a scale $M_X=4.13\times 10^{16}$ GeV, with the addition of extra Higgs components at an intermediate scale $M_{B-L}=8.54\times 10^{11}$ GeV. Here
$\alpha_{1}^{-1}=\frac{2}{5}\alpha_{(B-L)/2}^{-1}+\frac{3}{5}\alpha_{2}^{-1}$ at scale $M_{B-L}$.} \label{HWO2}
\end{center}
\end{figure}

In its minimal form, the SO(10) axion model includes the following Higgs contributions in the renormalization group equations (RGEs): the (1,1,3,+1) and (1,3,1,-1) components of the 126-plet, the (1,3,1,0) and (1,1,3,0) components of the 45-plet and the (1,2,2,0) components of the 10-plet are included between $M_X$ and $M_{B-L}$, while only the SM Higgs doublet in the (1,2,2,0) components of the 10-plet is included between $M_{B-L}$ and $M_Z$. The fermions in the 10-plet acquire masses $\sim M_{B-L}$. Using two loop RGEs for the standard model gauge couplings \cite{Machacek:1983tz}, we find $M_{B-L} = 3.5\times 10^{11}$ GeV and $M_X = 1.05\times 10^{15}$ GeV. 

To rectify the situation and obtain energy scales $\sim 10^{16}$ GeV, we consider threshold corrections from  suitable (intermediate mass $\sim M_{B-L}$) scalar components in the 45 and 126 multiplets. To illustrate this, we consider two appropriate combinations of these scalar components. In order to simplify matters, we include only the left-right symmetric scalar components in each of these cases. For our first example, we take (8, 1, 1, 0) from $45$, (3, 1, 1, $\sqrt{3/2}(-1/6)$) +   ($\overline{3}$, 1, 1, $\sqrt{3/2}(1/6)$) and $(3, 1, 3, \sqrt{3/2}(-1/6)$ + ($\overline{3}$, 3, 1, $\sqrt{3/2}(1/6)$) from $126$ all at scale $M_{B-L}$. Employing the two loop RGEs for the gauge couplings, we obtain unification at $M_X = 2.61\times 10^{16}$ GeV with $M_{B-L}=8.0\times 10^{11}$ GeV.

For our second example, we choose at scale $M_{B-L}$ the following multiplets: (8, 1, 1, 0) and (3, 1, 1, $\sqrt{3/2}(-2/6)$) +   ($\overline{3}$, 1, 1, $\sqrt{3/2}(2/6)$) from $45$, and (1, 2, 2, 0), $(3, 1, 3, \sqrt{3/2}(-1/6)$ + ($\overline{3}$, 3, 1, $\sqrt{3/2}(1/6)$) and (3, 2, 2, $\sqrt{3/2}(-2/6)$ + ($\overline{3}$, 2, 2, $\sqrt{3/2}(2/6)$) from $126$. In this case, we obtain unification at $M_X=4.13\times 10^{16}$ GeV with $M_{B-L}=8.54\times 10^{11}$ GeV (see Fig \ref{HWO2}).

\section*{Reheat Temperature and Non-Thermal Leptogenesis}

In SU(5), it is natural to consider the Yukawa coupling of the SU(5) singlet inflaton field $\phi$ to right-handed neutrinos $N_i$ $(i = 1,2,3)$. For simplicity we consider a simple Yukawa coupling
\begin{equation}
Y_N \, \phi  \, N \, N,
\end{equation}
where $Y_N \sim  M_N / \langle \phi \rangle $ is the coupling strength. For direct decay of the inflaton via this coupling, we take $2 M_N \lesssim m_\phi$; then, using $\langle \phi \rangle \sim$ (10 - 100) $ m_P$, we expect $Y_N \lesssim 10^{-6}$. Thus, we estimate the maximum value of the reheat temperature $T_r$ to be
\begin{equation}
T_r \sim \left(  \frac{m_\phi}{M} \right)   ( m_\phi \,  m_P )^{1/2}  \sim 10^{9} {\rm GeV}.
\end{equation}

In the SO(10) model, the value of $T_r$ is estimated from the coupling \cite{Holman:1982tb}
\begin{equation}
126 \times \overline{126} \times \phi^2,
\label{coupling1}
\end{equation}
which carries a negative sign in front in order to induce the intermediate scale breaking. (There is a similar coupling involving the 45-plet.) The coefficient of the expression in Eq. (\ref{coupling1}) is of order $(M_{B-L} /\langle \phi \rangle)^2$. Similarly as above, we take $2 M_{B-L} \lesssim m_\phi$. The decay rate of the inflaton into the scalar components ($\overline{10}$,1,3) of the 126-plet is then given by
\begin{equation}
\Gamma_{\phi} \sim \left( \frac{M_{B-L}}{M}\right) ^4 \frac{M^2}{m_\phi}.
\label{decayrate}
\end{equation}
The reheat temperature $T_r$ is estimated to be
\begin{equation}
T_r \sim \sqrt{\Gamma_{\phi}m_P} \sim \left( \frac{M_{B-L}}{M}\right) \left(\sqrt{\frac{m_P}{m_{\phi}}}\right)M_{B-L}.
\label{TR}
\end{equation}
The scalar bosons produced by the inflaton rapidly decay through their Yukawa couplings into right-handed neutrinos. Under the assumption of hierarchical right-handed neutrino masses, the lepton asymmetry is given by \cite{Fukugita:1986hr,Lazarides:1991wu,Lazarides:1996dv}
\begin{equation}
n_{L}/s\lesssim 3\times 10^{-10} \left( \frac{M_{N_i}}{M_{B-L}} \right) \left( \frac{T_{r}}{%
10^{6}{\rm~GeV}}\right) \left( \frac{m_{\nu_3}}{0.05{\rm~eV}}\right) ,  \label{10}
\end{equation}
where $M_{N_i}$ denotes the mass of the heaviest right-handed neutrino. From the experimental value of the baryon to photon ratio $\eta _{B}= 6.225\times 10^{-10}$ \cite{Dunkley:2008ie}, the required lepton asymmetry is found to be $n_{L}/s\approx 2.5\times10^{-10}$. With $T_r \sim  (10^6$-$10^8)$ GeV, these heavy neutrinos, with masses on the order of $8\times ( 10^{9}$-$10^{11})$ GeV or so, can give rise to the observed baryon asymmetry via non-thermal leptogenesis.

\section*{Magnetic Monopoles, Axion Domain Walls and Inflation}

The spontaneous breaking of any GUT symmetry yields topologically stable magnetic monopoles. In our SU(5) and SO(10) examples, these monopoles carry masses of order $\frac{M_{X}}{\alpha_{G}}$, where $\alpha_{G}\sim \frac{1}{30}$-$\frac{1}{40}$ denotes the GUT coupling constant. In the SU(5) model, for instance, monopoles are produced during inflation once the magnitude of the coefficient $\lambda_{2}\langle\phi\rangle^{2}$ associated with the \tr$\Sigma^{2}$ term in Eq. (\ref{VSU5}) exceeds $\sim H^{2}$, where $H = \sqrt{\frac{V_0}{3m_P^2}}$ ( $\sim (0.01-4) \times 10^{14}$ GeV ) denotes the Hubble constant during CW inflation. Using Table \ref{tableI} and Eq. (\ref{CWA}), it is readily checked that this occurs for values of $\phi\ll\phi_{0}$. Thus, the corresponding number of e-foldings is much greater than 50-60 and the monopoles are inflated away. This holds for the inflaton field rolling to its minimum from values smaller than its present VEV $M$. Clearly, the same results hold if the inflaton rolls to its minimum from values that are larger than $M$.

The breaking of U(1)$_{PQ}$ symmetry occurs at intermediate scales $\sim 10^{12}$ GeV $\ll H$. In this case the corresponding phase transition is completed at the end of inflation, with $T_{r}\sim 10^{6}$-$10^{9}$ GeV. Thus, it is prudent to ensure that there are no stable domain walls in the model by including in the SO(10) model two 10-plets of new matter fields (see Eq. (\ref{10plet})).

\section*{GUT Inflation and Proton Decay}

Our analysis above based on GUT inflation shows that the vacuum energy during inflation, for both the
CW and the Higgs potential, is of order $10^{16}$ GeV if we wish to obtain a scalar
spectral index close to 0.96, as suggested by WMAP5. In the CW case,
the SU(5) and SO(10) superheavy gauge bosons which mediate proton decay, are then expected to possess a mass $M_X$
which is a factor 2-4 larger than this scale. For the Higgs potential, the connection between the estimated
scale of vacuum energy and $M_X$ is somewhat tenuous, but it seems reasonable to assert that they are
of comparable magnitude. Put differently, the value for $M_X$ inferred above is a factor $\sim$ 1-10
larger than the lower bound $\sim 4 \times 10^{15}$ GeV obtained from the observed stability of the proton \cite{Nath:2006ut}. Thus, we expect proton lifetime estimates based on gauge boson mediated decays to be more in line with what one finds in supersymmetric GUTs, where $M_X \sim 2 \times 10^{16}$ GeV \cite{Raby:2008gh}. A few
lifetime estimates based on the naive expression for the decay rate \cite{Nath:2006ut}
\begin{equation}
\Gamma_p \approx \alpha_G^2 \, \frac{m_p^5}{M_X^4},
\end{equation}
are presented in Table \ref{tableIII}. (Here $m_p$ is the proton mass.)

The presence of an intermediate scale $ \sim 10^{12}$ GeV in the  SO(10) axion model  leads to the appearance of intermediate mass scalar leptoquarks $(3, 1,-2/3)$ which mediate proton decay. In this case, the proton decay rate is given by \cite{Nath:2006ut}
\begin{equation}
\Gamma_p \approx \vert Y_uY_d \vert^2 \, \frac{m_p^5}{M_{B-L}^4},
\end{equation}
where $Y_{u,d}$ denote the Yukawa couplings of the $u$ and $d$ quarks respectively. For an intermediate scale $\sim (0.5$-$1) \times 10^{12}$ GeV as in Fig. \ref{HWO2}, the proton lifetime is estimated to be of order $10^{34}$-$10^{35}$ years. 

\begin{table}[t]
{\centering
\begin{tabular}{||c|c||c|c||}
\hline 
\multicolumn{2}{||c||}{Coleman-Weinberg Potential} & \multicolumn{2}{c||} {Higgs Potential} \\
\hline  \hline  
$M_X \sim 2\, V^{1/4}_0$(GeV) & $\tau (p \rightarrow \pi^0 e^+ )$ (years) &  $M_X \sim  V^{1/4}_0$(GeV) & $\tau (p \rightarrow \pi^0 e^+ )$ (years) \\
\hline\hline
$ 5.0\times 10^{15}$ & $ 1.8\times 10^{34}$  &  $1.0\times 10^{16}$ & $ 2.8\times 10^{35}$ \\
\hline
$ 1.0\times 10^{16}$ & $ 2.8\times 10^{35}$ & $1.2\times 10^{16}$ & $5.8\times 10^{35}$ \\
\hline
$ 1.2\times 10^{16}$ & $5.8\times 10^{35}$ & $1.4\times 10^{16}$ & $1.1\times 10^{36}$ \\
\hline
$ 1.8\times 10^{16}$ & $2.9\times 10^{36}$ & $1.6\times 10^{16}$ & $1.8\times 10^{36}$  \\
\hline
$ 2.2\times 10^{16}$ & $6.6\times 10^{36}$ & $1.8\times 10^{16}$ & $2.9\times 10^{36}$ \\
\hline
$ 2.7\times 10^{16}$ & $1.5\times 10^{37}$ &  $2.1\times 10^{16}$ & $5.5\times 10^{36}$ \\
\hline
$ 3.5\times 10^{16}$ & $4.2\times 10^{37}$ &  $2.4\times 10^{16}$ & $9.3\times 10^{36}$ \\
\hline
$ 6.0\times 10^{16}$ & $3.6\times 10^{38}$ & $2.9\times 10^{16}$ & $2.0\times 10^{37}$ \\
\hline
\end{tabular} 
\caption{Superheavy gauge bosons masses and corresponding proton lifetimes with $\alpha_G = \frac{1}{35}$ in the CW and Higgs models. Note that since the lifetime depends only on $M_X$, the results shown here apply equally well to the BV and AV branches in each model.}\label{tableIII}
\centering}
\end{table}

\section*{Summary}

Assuming that inflation is associated with a GUT symmetry breaking phase transition (so that primordial GUT monopoles do not pose a cosmological problem), we have provided estimates for the energy scale of observable inflation which yield values for the scalar spectral index centered close to 0.96. For both CW and Higgs inflationary potentials, an estimate for the masses of the superheavy gauge bosons which mediate proton decay can then be derived. Masses of order $10^{16}$ GeV for these bosons are favored, yielding proton lifetimes of order $10^{34}$-$10^{38}$ years. In some SO(10) models with axion dark matter, lifetimes of order $10^{34}$-$10^{35}$ years are realized with proton decay being mediated by scalar leptoquarks possessing intermediate scale masses of order $10^{12}$ GeV.

\section*{Acknowledgments}
We thank Ilia Gogoladze and Nefer {\c S}eno$\breve{\textrm{g}}$uz for valuable discussions. This work is
supported in part by the DOE under grant \# DE-FG02-91ER40626 (Q.S. and M.R.),
by the Bartol Research Institute (M.R.), and by the NSF under grant \# DGE 0538555  (J.W.).



\begin{thebibliography}{99}

\expandafter\ifx\csname natexlab\endcsname\relax\def\natexlab#1{#1}\fi
\expandafter\ifx\csname bibnamefont\endcsname\relax
  \def\bibnamefont#1{#1}\fi
\expandafter\ifx\csname bibfnamefont\endcsname\relax
  \def\bibfnamefont#1{#1}\fi
\expandafter\ifx\csname citenamefont\endcsname\relax
  \def\citenamefont#1{#1}\fi
\expandafter\ifx\csname url\endcsname\relax
  \def\url#1{\texttt{#1}}\fi
\expandafter\ifx\csname urlprefix\endcsname\relax\def\urlprefix{URL }\fi
\providecommand{\bibinfo}[2]{#2}
\providecommand{\eprint}[2][]{\url{#2}}

\bibitem{Linde:2007fr}
  For a recent review and additional references, see
  A.~Linde,
  Lect.\ Notes Phys.\  {\bf 738}, 1 (2008)
  [arXiv:0705.0164 [hep-th]].
  See also
  D.~H.~Lyth and A.~Riotto,
  Phys.\ Rept.\  {\bf 314}, 1 (1999)
  [arXiv:hep-ph/9807278];
  G.~Lazarides,
  Lect.\ Notes Phys.\  {\bf 592}, 351 (2002)
  [arXiv:hep-ph/0111328].

\bibitem{Shafi:1983bd}
  Q.~Shafi and A.~Vilenkin,
  Phys.\ Rev.\ Lett.\  {\bf 52}, 691 (1984).

\bibitem{Pi:1984pv}
  S.~Y.~Pi,
  Phys.\ Rev.\ Lett.\  {\bf 52}, 1725 (1984);
  Q.~Shafi and A.~Vilenkin,
  Phys.\ Rev.\  D {\bf 29}, 1870 (1984).

\bibitem{Lazarides:1984pq}
  G.~Lazarides and Q.~Shafi,
  Phys.\ Lett.\  B {\bf 148}, 35 (1984).

\bibitem{Komatsu:2008hk}
  E.~Komatsu {\it et al.}  [WMAP Collaboration],
  arXiv:0803.0547 [astro-ph].

\bibitem{Shafi:2006cs}
  Q.~Shafi and V.~N.~{\c S}eno$\breve{\textrm{g}}$uz,
  Phys.\ Rev.\  D {\bf 73}, 127301 (2006)
  [arXiv:astro-ph/0603830].

\bibitem{Shiozawa}
  M. Shiozawa, plenary lecture given at SUSY 2008, Seoul, S. Korea.

\bibitem{Nath:2006ut}
  P.~Nath and P.~Fileviez Perez,
  Phys.\ Rept.\  {\bf 441}, 191 (2007)
  [arXiv:hep-ph/0601023] and references therein.

\bibitem{Kallosh:2007wm}
  R.~Kallosh and A.~Linde,
  JCAP {\bf 0704}, 017 (2007)
  [arXiv:0704.0647 [hep-th]].


\bibitem{Linde:2005ht}
\bibinfo{author}{\bibfnamefont{A.~D.} \bibnamefont{Linde}},
{\em Particle physics and inflationary cosmology} (\bibinfo{publisher}{Harwood Academic},
  \bibinfo{year}{1990}), hep-th/0503203.

\bibitem{Dvali:1994ms}
  G.~R.~Dvali, Q.~Shafi and R.~K.~Schaefer,
  Phys.\ Rev.\ Lett.\  {\bf 73}, 1886 (1994)
  [arXiv:hep-ph/9406319];
  E.~J.~Copeland, A.~R.~Liddle, D.~H.~Lyth, E.~D.~Stewart and D.~Wands,
  Phys.\ Rev.\  D {\bf 49}, 6410 (1994)
  [arXiv:astro-ph/9401011];
  M.~ur Rehman, V.~N.~Senoguz and Q.~Shafi,
  Phys.\ Rev.\  D {\bf 75}, 043522 (2007)
  [arXiv:hep-ph/0612023] and references therein.

\bibitem{Linde:1981mu}
  A.~D.~Linde,
  Phys.\ Lett.\  B {\bf 108}, 389 (1982);
  A.~Albrecht and P.~J.~Steinhardt,
  Phys.\ Rev.\ Lett.\  {\bf 48}, 1220 (1982).

\bibitem{Linde:1983gd}
  A.~D.~Linde,
  Phys.\ Lett.\ {\bf B129}, 177 (1983).


\bibitem{Liddle:1992wi}
  A.~R.~Liddle and D.~H.~Lyth,
  Phys.\ Lett.\ B {\bf 291}, 391 (1992),
  astro-ph/9208007;
  A.~R.~Liddle and D.~H.~Lyth,
  Phys.\ Rept.\  {\bf 231}, 1 (1993),
  astro-ph/9303019.

\bibitem{Salopek:1990jq}
  D.~S.~Salopek and J.~R.~Bond,
  Phys.\ Rev.\  D {\bf 42}, 3936 (1990).

\bibitem{Stewart:1993bc}
  E.~D.~Stewart and D.~H.~Lyth,
  Phys.\ Lett.\  B {\bf 302}, 171 (1993)
  [arXiv:gr-qc/9302019].

\bibitem{Destri:2007pv}
  C.~Destri, H.~J.~de Vega and N.~G.~Sanchez,
  Phys.\ Rev.\  D {\bf 77}, 043509 (2008)
  [arXiv:astro-ph/0703417].

\bibitem{Smith:2008pf}
  T.~L.~Smith, M.~Kamionkowski and A.~Cooray,
  arXiv:0802.1530 [astro-ph].



\bibitem{NeferSenoguz:2008nn}
  V.~N.~{\c S}eno$\breve{\textrm{g}}$uz, and Q.~Shafi,
  Phys.\ Lett.\  B {\bf 668}, 6 (2008)
  [arXiv:0806.2798 [hep-ph]].

\bibitem{Dorsner:2005fq}
  For a recent discussion see
I.~Dorsner and P.~Fileviez Perez,
  Nucl.\ Phys.\  B {\bf 723}, 53 (2005)
  [arXiv:hep-ph/0504276], and references therein.


\bibitem{Chkareuli:1994ng}
  J.~L.~Chkareuli, I.~G.~Gogoladze and A.~B.~Kobakhidze,
  Phys.\ Lett.\  B {\bf 340}, 63 (1994).

\bibitem{Holman:1982tb}
  R.~Holman, G.~Lazarides and Q.~Shafi,
  Phys.\ Rev.\  D {\bf 27}, 995 (1983).


\bibitem{Fukugita:1986hr}
\bibinfo{author}{\bibfnamefont{M.}~\bibnamefont{Fukugita}} \bibnamefont{and}
  \bibinfo{author}{\bibfnamefont{T.}~\bibnamefont{Yanagida}},
  \bibinfo{journal}{Phys. Lett.} \textbf{\bibinfo{volume}{B174}},
  \bibinfo{pages}{45} (\bibinfo{year}{1986}).


\bibitem{Lazarides:1991wu}
For non-thermal leptogenesis see
\bibinfo{author}{\bibfnamefont{G.}~\bibnamefont{Lazarides}} \bibnamefont{and}
  \bibinfo{author}{\bibfnamefont{Q.}~\bibnamefont{Shafi}},
  \bibinfo{journal}{Phys. Lett.} \textbf{\bibinfo{volume}{B258}},
  \bibinfo{pages}{305} (\bibinfo{year}{1991}).


\bibitem{Pati:1974yy}
  J.~C.~Pati and A.~Salam,
  Phys.\ Rev.\  D {\bf 10}, 275 (1974)
  [Erratum-ibid.\  D {\bf 11}, 703 (1975)];
  G.~Senjanovic and R.~N.~Mohapatra,
  Phys.\ Rev.\  D {\bf 12}, 1502 (1975);
  R.~N.~Mohapatra and J.~C.~Pati,
  Phys.\ Rev.\  D {\bf 11}, 566 (1975);
  G.~Senjanovic,
  Nucl.\ Phys.\  B {\bf 153}, 334 (1979);
  M.~Magg, Q.~Shafi and C.~Wetterich,
  Phys.\ Lett.\  B {\bf 87}, 227 (1979);
  M.~Cvetic,
  Nucl.\ Phys.\  B {\bf 233}, 387 (1984).

\bibitem{Preskill:1982cy}
  J.~Preskill, M.~B.~Wise and F.~Wilczek,
  Phys.\ Lett.\  B {\bf 120}, 127 (1983);
  L.~F.~Abbott and P.~Sikivie,
  Phys.\ Lett.\  B {\bf 120}, 133 (1983);
  M.~Dine and W.~Fischler,
  Phys.\ Lett.\  B {\bf 120}, 137 (1983).


\bibitem{Sikivie:1982qv}
 P.~Sikivie,
 Phys.\ Rev.\ Lett.\  {\bf 48}, 1156 (1982).


\bibitem{Lazarides:1982tw}
 G.~Lazarides and Q.~Shafi,
 Phys.\ Lett.\  B {\bf 115}, 21 (1982).

\bibitem{Machacek:1983tz}
  M.~E.~Machacek and M.~T.~Vaughn,
  Nucl.\ Phys.\  B {\bf 222}, 83 (1983);
  Nucl.\ Phys.\  B {\bf 236}, 221 (1984);
  Nucl.\ Phys.\  B {\bf 249}, 70 (1985);
  V.~D.~Barger, M.~S.~Berger and P.~Ohmann,
  Phys.\ Rev.\  D {\bf 47}, 1093 (1993)
  [arXiv:hep-ph/9209232].


\bibitem{Lazarides:1996dv}
  G.~Lazarides, R.~K.~Schaefer and Q.~Shafi,
  Phys.\ Rev.\  D {\bf 56}, 1324 (1997)
  [arXiv:hep-ph/9608256];
\bibinfo{author}{\bibfnamefont{T.}~\bibnamefont{Asaka}},
  \bibinfo{author}{\bibfnamefont{K.}~\bibnamefont{Hamaguchi}},
  \bibinfo{author}{\bibfnamefont{M.}~\bibnamefont{Kawasaki}}, \bibnamefont{and}
  \bibinfo{author}{\bibfnamefont{T.}~\bibnamefont{Yanagida}},
  \bibinfo{journal}{Phys. Rev.} \textbf{\bibinfo{volume}{D61}},
  \bibinfo{pages}{083512} (\bibinfo{year}{2000}), hep-ph/9907559;
\bibinfo{author}{\bibfnamefont{V.~N.} \bibnamefont{{\c
  S}eno$\breve{\textrm{g}}$uz}} \bibnamefont{and}
  \bibinfo{author}{\bibfnamefont{Q.}~\bibnamefont{Shafi}},
  \bibinfo{journal}{Phys. Rev.} \textbf{\bibinfo{volume}{D71}},
  \bibinfo{pages}{043514} (\bibinfo{year}{2005}), hep-ph/0412102.

\bibitem{Dunkley:2008ie}
  J.~Dunkley {\it et al.}  [WMAP Collaboration],
  arXiv:0803.0586 [astro-ph].

\bibitem{Raby:2008gh}
  For a recent review and additional references see  
  S.~Raby,
  arXiv:0807.4921 [hep-ph].

\end{thebibliography}
\end{document}